\documentstyle[11pt,epsfig]{article} 

\newcommand{\NPB}[3]{{\it Nucl.\ Phys.\/} {\bf B#1}, {#2} (19{#3})}  
\newcommand{\PLB}[3]{{\it Phys.\ Lett.\/} {\bf #1B}, {#2} (19{#3})} 
\newcommand{\PRD}[3]{{\it Phys.\ Rev.\/}  {\bf D#1}, {#2} (19{#3})} 
 
\newcommand{\PRL}[3]{{\it Phys.\ Rev.\ Lett.\/} {\bf #1}, {#2} (19{#3})}

\def\n55{n_{5\overline 5}}

\def\QQa{\renewcommand{\baselinestretch}{1.2}\Huge\large\normalsize}

\begin{document}

\begin{flushright}
UG-FT-88/98\\
May 1998
\end{flushright}

\vspace*{3.2cm}

\begin{center}
\begin{bf}
\centerline {\large \bf LIMITS ON THE MASS OF THE LIGHTEST}
\centerline {\large  \bf         SUSY HIGGS               }
\end{bf}
\vspace*{0.7cm}

M. Masip and R. Mu{\~n}oz-Tapia\footnote{Talk presented 
at the XXXIIIrd Rencontres de Moriond, Electroweak and Unified Theories}\\
{\it Departamento de F{\'\i}sica Te{\'o}rica y del Cosmos,\\
Universidad de Granada,
18071 Granada, Spain}

\vspace{2cm}

\parbox[t]{12cm}{
We study the limits on the mass of the lightest Higgs 
in supersymmetric models extended with a gauge singlet
when perturbative unification is required. We find that
when maximum intermediate matter is added, the different 
evolution of the gauge couplings raises the mass bound 
from 135 GeV to 155 GeV. In these models perturbative 
unification of the gauge couplings is achieved in a 
natural way.
 }
\end{center}

\bigskip
\bigskip
\QQa

The minimal supersymmetric extension of the
standard model (MSSM) successfully
accommodates the standard model (SM) at low energies, 
and is arguably the best 
candidate to describe physics at energies beyond 1 TeV. The MSSM also 
predicts the unification of the
gauge couplings at a scale of $M_X\sim 2\times 10^{16}$ GeV,
a striking result that spans over many orders of
magnitude. This unification can be regarded as a high energy prediction 
obtained from low energy measurements \cite{unification}.

Supersymmetric models have been up to now flexible enough to respect all 
experimental constraints. However, this flexibility does not 
translate into a complete lack of predictivity. The most compelling
low energy outcome is probably the
existence of a light Higgs. In the MSSM
the lightest state is a CP even neutral Higgs. Its mass is 
bounded at tree level by
 \begin{equation}
m_h^2 \le M_Z^2\;\cos^22\beta\;, 
\label{mh0}
\end{equation}
where  $\tan\beta$	 is
the ratio between the 
vacuum expectation values 
$v$ and $\bar v$ of the Higgs fields $H$ and $\bar H$ that give masses
to the up and down quarks respectively \cite{hhg}. Eq.~1 is a very stringent
bound as the only supersymmetric parameter that appears is
$\tan\beta$. It tells us that at tree level the lightest Higgs
should have a mass lighter than $M_Z$.
Radiative corrections to Eq.~1 can be 
calculated 
and are known to relax the bound considerably \cite{radiative1}.

\begin{figure}
\centering
\begin{minipage}[t]{130mm}
\mbox{%
\epsfig{file=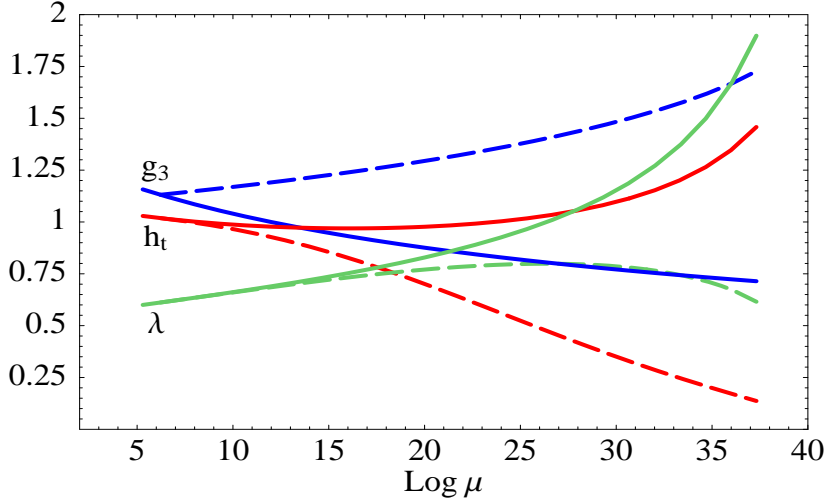 
       ,height=7cm  
       ,width=12cm   
       }%
       }
\caption{Evolution of $\lambda$, $h_t$ and $g_3$ with (dashed lines)
 four $5+\bar{5}$ SU(5) representations at 1 TeV and without (solid lines)
 extra matter} 
\label{fig1}
\end{minipage}   
\end{figure}
   
When the MSSM is enlarged with gauge singlets, trilinear terms
appear in the superpotential
\begin{equation}
W \supset \lambda\; SH\bar H 
\label{p1}
\end{equation}
and the tree-level bound becomes
\begin{equation}
m_h^2 \le M_Z^2\;\cos^22\beta + \lambda^2 \nu^2\;\sin^22\beta \;, 
\label{mh1}
\end{equation}
with $\nu=\sqrt{v^2+\bar v^2}=174$ GeV.
For an unrestricted value of $\lambda$ no bound can be 
extracted from (\ref{mh1}).
However, since the celebrated unification of the gauge couplings
is obtained perturbatively, it is physically sound to impose
the perturbative unification criterion. That is, all coupling
constants should remain perturbative up to the unification
scale \cite{eq92}. This condition translates into a bound on $\lambda$.
Indeed, consider the beta-function governing the evolution of
$\lambda$ at one loop
\begin{equation}
\beta_\lambda = {\lambda\over 16 \pi^2} (
4 \lambda^2 + 3 h_t^2 + 3 h_b^2 - g_1^2 - 3 g_2^2)\;, 
\label{beta}
\end{equation}
where $h_t$ and $h_b$ are the top and bottom Yukawa couplings,
and $g_1$ and $g_2$ are the $U(1)_Y$ and $SU(2)_L$ gauge
couplings, respectively. The positive coefficients of $\lambda$
and $h_t$ (the other terms have a smaller impact)
imply that the running value of  $\lambda$ increases
with energy. The starting low energy value of $\lambda$
 has to be small enough to avoid
blowing up before reaching the unification
scale. For a top quark of 180 GeV this argument implies a bound
on $\lambda$ of $\sim$ 0.7 \cite{kw}.

A possible way to increase the allowed  low energy value of
$\lambda$ is to introduce matter fields at intermediate scales.
The effect of these fields would be indirect in the sense that 
they will modify the evolution of the gauge couplings. This 
argument was outlined in \cite{kane}. There, Higgs 
doublets were considered because they increase the evolution
rate of $g_1$ and $g_2$ couplings  that enter with negative sign in 
Eq.~(\ref{beta}). However, it was found that the effect was always
small, because the addition of matter is also limited by the perturbative
criterion for the gauge couplings. Furthermore, the presence
of such doublets spoils their unification.

\begin{figure}
\centering
\begin{minipage}[t]{130mm}
\epsfig{file=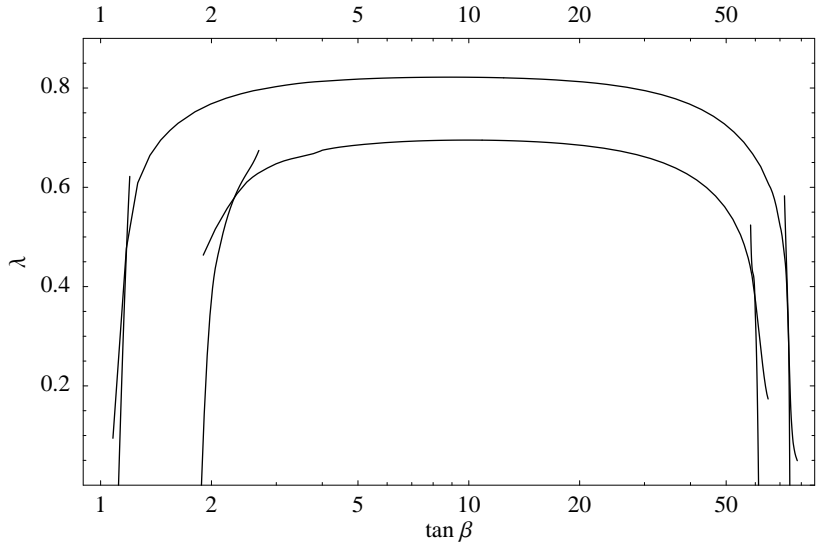 
        ,height=8cm  
        ,width=12cm   
       }
\caption{Limits on the value of $\lambda$ at the weak scale.
We plot the singlet model in the cases with
a maximal matter content  at intermediate scales
(upper) and without extra matter (lower).} 
\label{fig2} 
\end{minipage}
 \end{figure}

This analysis can nevertheless be improved \cite{mmp}. 
First, it is important
to note that  $g_3$ and $h_t$, have the highest numerical value and
therefore have a significant impact in the evolution equations. If the 
evolution rate of $g_3$ is increased (by the addition of coloured matter),
then the evolution rate of $h_t$ is slowed because $g_3$ enters with
a large negative coefficient in its beta--function (see \cite{mmp}
for the complete renormalization group equations up to two loops).
Lower values of $h_t$ in Eq.~(\ref{beta}) imply higher allowed 
starting values for $\lambda$. This effect is shown in Fig.~1, 
where the evolution of $g_3$, $h_t$ and $\lambda$ is depicted
with and without intermediate matter. One sees that when matter
is present $h_t$ even decreases with energy, mainly due to the
higher evolution rate of $g_3$.
Second, there is a way to introduce extra matter fields that respect
perturbative unification of the gauge couplings. Complete
representations of a simple
group [$SU(5)$, $SO(10)$, $E_6$,...] that contains
$SU(3)_C\times SU(2)_L\times U(1)_Y$ 
as a subgroup modify the 
(one-loop) 
running of the three gauge couplings in such way that 
they still meet at the same unification scale $M_X$,
but with a higher final value.
This scenario allows larger intermediate values of $g_1$ and $g_2$
together with smaller values of $h_t$ and defines the setting for the
absolute perturbative bound on $\lambda$. The presence of extra matter 
can be motivated by models with gauge
mediated supersymmetry breaking (GMSB) \cite{gr}. Minimal scenarios of GMSB
could in fact be closer to the singlet model than to the MSSM.

We can now obtain the bound on $\lambda$ with vector--like matter
at intermediate scales. We consider the full
two loop renormalization group equations for the evolution of the 
couplings. The first and second coefficients of the 
beta--functions are of opposite sign. The couplings would
then evolve up to the point where the beta--functions cancel. 
We shall consider that any coupling $g$
is non-perturbative if at a scale below $M_X$ the running value 
is ${g\over 4\pi} > 0.3$. The change from the
perturbative to the non-perturbative regime 
(when the constants reach their limiting value) is quite abrupt, and 
therefore the results do not depend appreciably on the actual cutting value
chosen.

The results for the bounds on $\lambda$ are depicted in Fig.~2.
The lower curve corresponds to the
case of the singlet model without extra matter. On the left and right
hand side of the plot the first couplings to become non--perturbative
are the top and bottom Yukawas respectively. In the middle zone,
the $\lambda$ coupling itself imposes the dominant condition. The allowed
range for $\tan \beta$ is $ 1.88 \leq \tan \beta \leq 51.2$. Beyond these
values, even for $\lambda=0$, the perturbative criterion is not satisfied.
The absolute limit is $\lambda=0.69$, which occurs at $\tan \beta=10$.

In the upper curve a maximal content of extra matter is added before the
gauge couplings become non--perturbative. There are several ways to
reach this maximal content. We have plotted the case of four 
 5+$\bar{5}$ representations
of SU(5), which is the lowest dimensional vector representation of 
a single group containing the SM symmetries, introduced
at 250 GeV. The same three zones appear, but now the allowed range of 
$\tan \beta$ is enlarged, $\tan \beta$ is $ 1.19 \leq \tan \beta \leq 74.4$.
This a relevant outcome, as in these models the maximum value of the
mass bound is obtained at lower $\tan \beta$. The absolute maximum on 
$\lambda$ is 0.82 at $\tan \beta=8$, a 19\% higher than the previous
case.

\begin{figure}
\centering
\begin{minipage}[t]{130mm}
\epsfig{file=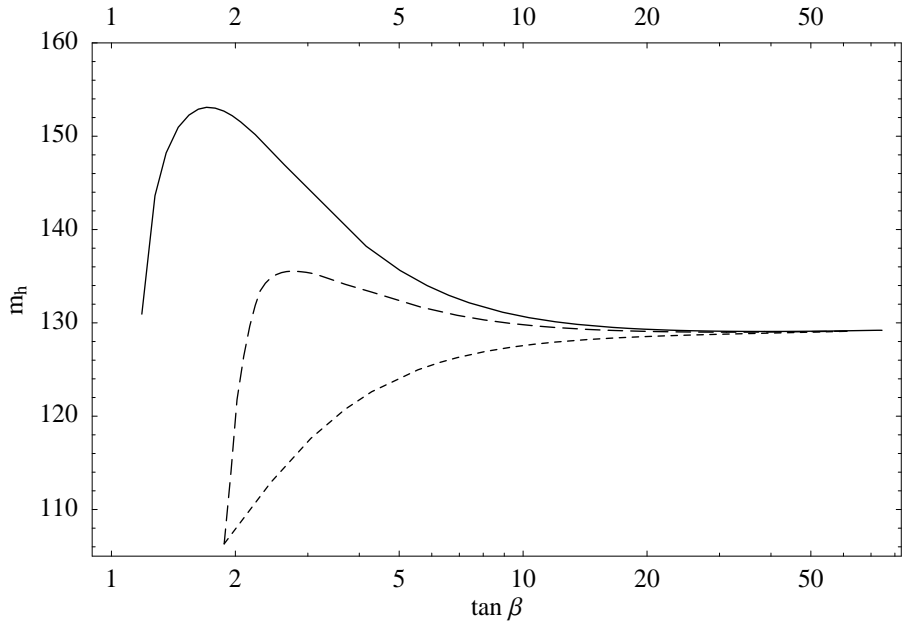 
       ,height=9cm  
        ,width=13cm}
\caption{Limit on $m_h$ (in GeV) in the MSSM, in the singlet model with
no intermediate matter, and in the singlet model with a maximal
matter content at intermediate scales. 
} 
\label{fig3}
 \end{minipage}        
\end{figure}
The bounds on $\lambda$ can now be translated into mass bounds, once the
radiative corrections of the top-quark loops are added \cite{radiative2}, 
which
amount a shift of around 30 GeV for squark masses of 1 TeV and no mixing. 
Our results are shown in Fig.~3. As is clear from the plot, if the
presence of a singlet takes the MSSM bound from 
$m_h\le 128$ GeV to $m_h\le 135$ GeV,
the presence of matter at intermediate scales pushes
this bound further up, to $m_h\le 155$ GeV. The actual numerical values
of the bounds depend on the precise input parameters (top mass, strong
coupling constant, mixings, etc.), however it is a generic feature that
the addition of vector--like matter has a significant impact in raising
the  value of the bound without spoiling unification of the gauge
couplings.

\section*{Acknowledgments} 

This work was supported by CICYT under contract 
AEN96-1672 and by the Junta de Andaluc{\'\i}a under contract
FQM-101.

\end{document}